\documentclass[aps,prl,twocolumn,superscriptaddress,showpacs,footnote]{revtex4}

\usepackage{graphicx,pstricks}
\def\lsim{\stackrel{\scriptstyle <}{\phantom{}_{\sim}}}
\def\gsim{\stackrel{\scriptstyle >}{\phantom{}_{\sim}}}
\newcommand{\be}{\begin{equation}}
\newcommand{\ee}{\end{equation}}
\newcommand{\ba}{\begin{eqnarray}}
\newcommand{\ea}{\end{eqnarray}}

\begin{document}

\title{On the Cooling of the Neutron Star in Cassiopeia A}

  \author{D. Blaschke} \affiliation{Institute for Theoretical Physics,
    University of Wroc{\l}aw, 50-204 Wroc{\l}aw, Poland}
  \affiliation{Bogoliubov Laboratory for Theoretical Physics, Joint
    Institute for Nuclear Research, 141980 Dubna, Russia}

  \author{H. Grigorian} \affiliation{Department of Theoretical
    Physics, Yerevan State University, 
    375025 Yerevan, Armenia}

  \author{D. N. Voskresensky} \affiliation{ National Research Nuclear
    University (MEPhI), 115409 Moscow, Russia} \affiliation{ExtreMe
    Matter Institute EMMI and Research Division, GSI Helmholtzzentrum
    f\"ur Schwerionenforschung, Planckstra${\beta}$e 1, 64291
    Darmstadt, Germany}

  \author{F. Weber} \affiliation{Department of Physics, San Diego
    State University,
    San Diego, California 92182, USA}

\begin{abstract}
  We demonstrate that the high-quality cooling data observed for the
 young neutron star in the supernova remnant Cassiopeia A over the
  past 10 years--as well as all other reliably known temperature data
 of neutron stars--can be comfortably explained within the "nuclear
 medium cooling" scenario. The cooling rates of this scenario account
 for medium-modified one-pion exchange in dense matter and
 polarization effects in the pair-breaking formations of superfluid
 neutrons and protons.
 Crucial for the successful description of the observed data is
 a substantial reduction of the thermal conductivity,
 resulting from a suppression of both the electron and nucleon contributions
 to it by medium effects.
  We also find that possibly in as little as about ten years of
 continued observation, 
 the data may tell whether or not
 fast cooling processes are active in this neutron star. 
\end{abstract}



\maketitle

{\it Introduction.--}
The isolated neutron star in Cassiopeia~A (Cas~A) was discovered in
1999 by the {\em Chandra} satellite \cite{Tananbaum:1999kx}.  Its
association with the historical supernova SN~1680
\cite{Ashworth:1980vn} gives Cas~A an age of 330~years, in agreement
with the nebula's kinematic age \cite{Fesen:2006ys}.  The distance to
the stellar remnant is estimated to be $3.4^{+0.3}_{-0.1}$~kpc
\cite{Reed:1995zr}.  The thermal soft X-ray spectrum of Cas~A can be
fitted with a non-magnetized carbon atmosphere model, a surface
temperature of $2\times 10^6$~K, and an emitting radius of 8 to 17~km
\cite{Ho:2009fk}.  Analyzing the data from 2000 to 2009, Heinke \& Ho
\cite{Heinke:2010xy} reported a rapid decrease of Cas~A's surface
temperature over this 10-year period, from $2.12 \times 10^6$ to $2.04
\times 10^6$~K.  Such a rapid drop in temperature conflicts with
standard cooling scenarios based on the efficient modified Urca (MU)
process ~\cite{Yakovlev:2000jp,Page:2006ly}.  First interpretations of
Cas~A's temperature data were provided very recently by Page {\it et
  al.} \cite{Page:2010aw} and Yakovlev {\it et al.}
\cite{Shternin:2010qi,Yakovlev:2010ed}.

The interpretation of Page {\it et al.} \cite{Page:2010aw} is based on
the ``minimal cooling'' paradigm \cite{Page:2004fy}, where a minimal
number of cooling processes is taken into account. These are photon
emission, the MU process, nucleon-nucleon ($NN$) bremsstrahlung (NB)
and the neutron ($n$) and proton ($p$) pair breaking-formation
processes (nPBF and pPBF). The latter are particularly important in
the ultra-dense cores of neutron stars
\cite{Flowers:1976vn,Voskresensky:1987hm,Schaab:1996gd}, where
neutrons form Cooper pairs in the $^3P_2$ channel and proton pairing
occurs in the $^1S_0$ channel.  To calculate the $NN$ interaction
entering the emissivities of the MU and NB processes the minimal
cooling scenario employs the free one-pion exchange (FOPE) model
\cite{Friman:1978zq}.  As shown in \cite{Page:2010aw}, the Cas~A data
can be neatly reproduced by assuming a large value for the proton
pairing gap throughout the entire stellar core and by fixing the
critical temperature for the neutron $^3P_2$ pairing gap at around
$0.5\times 10^9$~K.  The result is mildly sensitive to the neutron 
star mass.
  Surface temperature--age data of other neutron stars,
which do not lie on the cooling curve of Cas~A, are explained within
the minimal cooling scenario mainly by assuming variations in the
light element mass of the envelopes of these stars.

The work of Yakovlev {\it et al.}
\cite{Shternin:2010qi,Yakovlev:2010ed} includes all emission processes
which are part of the minimal cooling paradigm and uses also the FOPE
to model the $NN$ interaction. As in \cite{Page:2010aw}, it is assumed
that the proton gap is large and non-vanishing in the entire stellar
core.  The latter assumption facilitates a strong suppression of the
emissivity of the MU process.  The value and the density dependence of
the $^3P_2$ neutron gap are fitted to the Cas~A data, leading to a
critical temperature of $(0.7-0.9)\times 10^9$~K for the neutron
pairing gap. Both groups therefore came to the striking conclusion
that the temperature data of Cas~A allow one to extract the value of
the $^3P_2$ neutron pairing gap.

In this Letter, we present the ``nuclear medium cooling scenario'' as
an alternative model for the successful description of the temperature
data of Cas~A. Aside from describing the Cas~A data extremely well,
this model reproduces also all other presently known temperature data
of NSs, without the need of making any additional
assumptions. Before representing the stellar cooling results, we
outline the key features of the nuclear medium cooling scenario next.

\medskip

{\it Nuclear medium cooling.--}
  Motivated by the fact that the existing temperature--age data of 
neutron stars seem
  to be incompatible with a unique cooling evolution, the nuclear
medium cooling scenario has been worked out in
Refs.~\cite{Voskresensky:1986af,Voskresensky:1987hm,
Migdal:1990vm,Voskresensky:2001fd}.
It provides a microscopic justification for a strong
dependence of the main cooling mechanisms on the density
(and thus on the  neutron star mass).
The nuclear medium cooling scenario has been successfully applied to the
description of the body of known surface temperature--age data of neutron stars
\cite{Schaab:1996gd,Blaschke:2004vq,Grigorian:2005fn}. The scenario
addresses the often disregarded role of medium effects on the MU and
NB processes.
Furthermore, as it is commonly accepted, the neutron and proton
superfluidity with density dependent pairing gaps is causing an
exponential suppression of neutrino emissivities of the nucleon
processes and of the nucleon specific heat, and opens up the new class
of nPBF and pPBF processes.  We also want to stress that the thermal
conductivity is essential for the cooling of young objects such as Cas~A.
The next paragraph is devoted to a brief discussion of these
issues. For more details, we refer to
\cite{Migdal:1990vm,Voskresensky:2001fd,Blaschke:2004vq}.

{\it 1. Free versus medium-modified one-pion-exchange in dense
  matter:} The insufficiency of the FOPE model for the description of
the $NN$-interaction is a known issue \cite{Voskresensky:2001fd}.
Indeed, calculating the MU emissivity perturbatively one may use both
the Born $NN$ interaction amplitude given by the FOPE and the
imaginary part of the pion self-energy.  In the latter case one needs
to expand the exact pion Green's function $D_{\pi}(\omega,k)=[\omega^2
-m_{\pi}^2 -k^2 -\Pi(\omega,k,n)]^{-1}$ to second order using for the
polarization function $\Pi(\omega,k,n)$ the
perturbative one-loop diagram, $\Pi_0(\omega,k,n)$. For $k=k_0$, which
is the pion momentum at the minimum of the effective pion gap
$\omega^{*\,2}=-D_{\pi}^{-1}(\omega =0,k =k_0)$, the polarization
function $\Pi_0 (\omega,k=k_0\simeq p_{{\rm F},n},n)$ yields however a
strong $NN$ attraction.  (Here $m_{\pi}$ is the pion mass and $p_{{\rm
    F},n}$ is the neutron Fermi momentum.)  This attraction is so
strong that it would trigger a pion condensation instability already
at low baryon densities of $n \sim 0.3\, n_0$, which is in disagreement
with experimental data on atomic nuclei.  ($n_0 = 0.16$~fm$^{-3}$
denotes the nuclear saturation density.)  The discrepancy is resolved
by observing that together with a pion softening (i.e., a decrease of
the effective pion gap $\omega^{*}(n)$ with increasing density) one
needs to include the repulsion from the dressed $\pi NN$ vertices,
$\Gamma \simeq [1+C (n/n_0)^{1/3}]^{-1}$, with $C\simeq 1.4\div 1.6$.
A consistent description of the $NN$ interaction in matter should thus
use a medium modified one-pion exchange (MOPE) interaction
characterised by the full Green function of the dressed pion, dressed
vertices $\Gamma(n)$, and a residual $NN$ interaction, as done in
  this Letter.
According to  \cite{
Voskresensky:1986af,Migdal:1990vm},
the main contribution for $n>n_0$ is given by MOPE whereas the relative
contribution of the  residual interaction decreases with increasing density.
Following the model used in \cite{Blaschke:2004vq,Grigorian:2005fn}  pion
condensation may arise only for $n\geq n_{cr}^{\pi}= 3\, n_0$, i.e.,
for NS masses $M\geq 1.32 M_{\odot}$  within a relativistic version of
the APR equation of state
\cite{Akmal:1998qf}
which we use.  In the calculation of
the neutrino emissivity not only radiation from the nucleon legs but
also from intermediate reaction states is now allowed. With such an
interaction the ratio of the emissivity of the medium modified Urca
(MMU) to the MU process,
\be \frac{\epsilon_{\nu}[\rm MMU]}{\epsilon_{\nu}[\rm MU]}\sim
10^{3}\left(\frac{n}{n_0}\right)^{10/3}
\frac{\Gamma^6(n)}{[\omega^{*}(n)/m_{\pi}]^8} \, ,  \ee
strongly increases with density (for $n\gsim n_0$). Although an
increase of the ratio of emissivities of the medium modified nucleon
(neutron) bremsstrahlung process (MnB) to the unmodified
bremsstrahlung (nB) is less pronounced, the MnB process, being not
affected by the proton superconductivity, may yield a relatively large
contribution in the region of a strong proton pairing.

{\it 2. Pair-breaking formation:} The important role of polarization
effects in pPBF and nPBF processes was first noted in
\cite{Voskresensky:1987hm}.  Recently, additional support came from an
analysis of the vector current conservation in PBF reactions
\cite{Leinson:2006gf,Kolomeitsev:2008mc}.  In these reactions,
diagrams with the normal and anomalous Green functions turn out to
cancel each other, so that the main contribution to the PBF emissivity
comes from processes of the axial current \cite{Kolomeitsev:2008mc}.
Another important in-medium effect was recently observed in the
calculation of the neutron pairing gap $^3P_2$.  Taking into account
the polarization effects, it was shown in \cite{Schwenk:2003bc} that
the associated  gap $\Delta_{nn} (^3P_2)\lsim $~keV, i.e. it is dramatically 
suppressed compared to BCS based calculations \cite{Takatsuka:2004zq}.
For completeness, we also mention the possibility of a  strong enhancement
(more than 1 MeV) of the gap, as argued in \cite{Khodel:2004nt}.
At first glance, the results of \cite{Schwenk:2003bc} and \cite{Khodel:2004nt}
seem to illustrate uncertainties in value of the $^3P_2$ gap, which would
suggest to treat $\Delta_{nn} (^3P_2)$ as a free parameter in cooling studies.
This, however, is not the case since the solution of the gap equation
of \cite{Khodel:2004nt} exists only for $\Delta_{nn} (^3P_2)\gsim
1$~MeV and disappears for smaller values of the gap.  Moreover, they
use the approximation $0<\omega^{*\,2} (n)\ll m_\pi^2$ so that their
new solution may exist only within a narrow range of the critical
point for the onset of pion condensation.
In realistic treatments, pion condensation appears always as a first order phase
transition \cite{Migdal:1990vm} with a jump of $\omega^{*\,2}$ from a
positive to a negative value.  The required small values of the pion
gap may therefore not be achieved. Moreover, \cite{Grigorian:2005fn}
has verified that the cooling data are hardly described if the
gap $\Delta_{nn} (^3P_2)$ were large ($\gsim 1$~MeV) over a broad density
region.
We therefore disregard the possibility of a large value of $\Delta_{nn} (^3P_2)$
and adopt  tiny $\Delta_{nn} (^3P_2)$ following
\cite{Schwenk:2003bc}.
The $\Delta_{nn} (^1S_0)$ proton gap is taken from \cite{APW}.
Two different models, labeled I and II, are used for the $\Delta_{pp} (^1S_0)$
gap \cite{Blaschke:2004vq,Grigorian:2005fn}.  Model I is from
\cite{Yakovlev:2003qy} and model II is from the calculations in
\cite{Takatsuka:2004zq}. NS cooling data can be well
described within the nuclear medium cooling scenario for both models I
and II \cite{Blaschke:2004vq}, provided the $\Delta_{nn} (^3P_2)$ is
strongly suppressed in agreement with Schwenk \& Friman
\cite{Schwenk:2003bc}.

{\it 3. The heat conductivity:}
The heat conductivity, $\kappa$, of superfluid neutron star matter is another 
key ingredient crucial for the cooling
  of young neutron stars, such as Cas~A.  
It is given by $\kappa =\sum_i \kappa_i$ where $\kappa_i$ are the  partial
  contributions to $\kappa$.
 In \cite{Blaschke:2004vq} and
  \cite{Yakovlev:2003qy} the electron heat conductivity $\kappa_e$
  computed according to \cite{Baiko:2001cj} was used. More
  recent studies \cite{Shternin:2007ee} showed that for
  temperature and density regions where nucleons are non-superfluid,
  Landau damping reduces $\kappa_e$ below $\kappa_n$. Proton
  superfluidity causes a Meissner screening of transverse
  photons.
Taking all these effects into account reduces the
    total thermal conductivity \cite{Shternin:2007ee} by an order of
    magnitude.
Also, in \cite{Blaschke:2004vq} we argued that medium effects may significantly
suppress the neutron contribution $\kappa_n$ to the thermal conductivity, the 
effect being not included in \cite{Shternin:2007ee}. 
Indeed, $\kappa_n\propto 1/|M^2|$, where $M$ is the $NN$ interaction matrix
element discussed already in the MOPE context as being enhanced for
$n\gsim n_0$.
The impact of a low thermal conductivity on the thermal evolution of
neutron stars accomplished  by introducing  a factor
$\zeta_\kappa =0.3$ was demonstrated in  Fig.~17 of \cite{Blaschke:2004vq}.
The net effect is a delay of the temperature decline of young ($\sim 300$~yr) 
neutron stars.
This idea of a possible strong suppression of the thermal conductivity, as
supported by
\cite{Shternin:2007ee}, 
proves essential for the explanation of the rapid cooling of Cas~A in this
letter.

\medskip

\begin{figure}[tbh]
   \includegraphics[width=0.45\textwidth,height=0.45\textwidth]{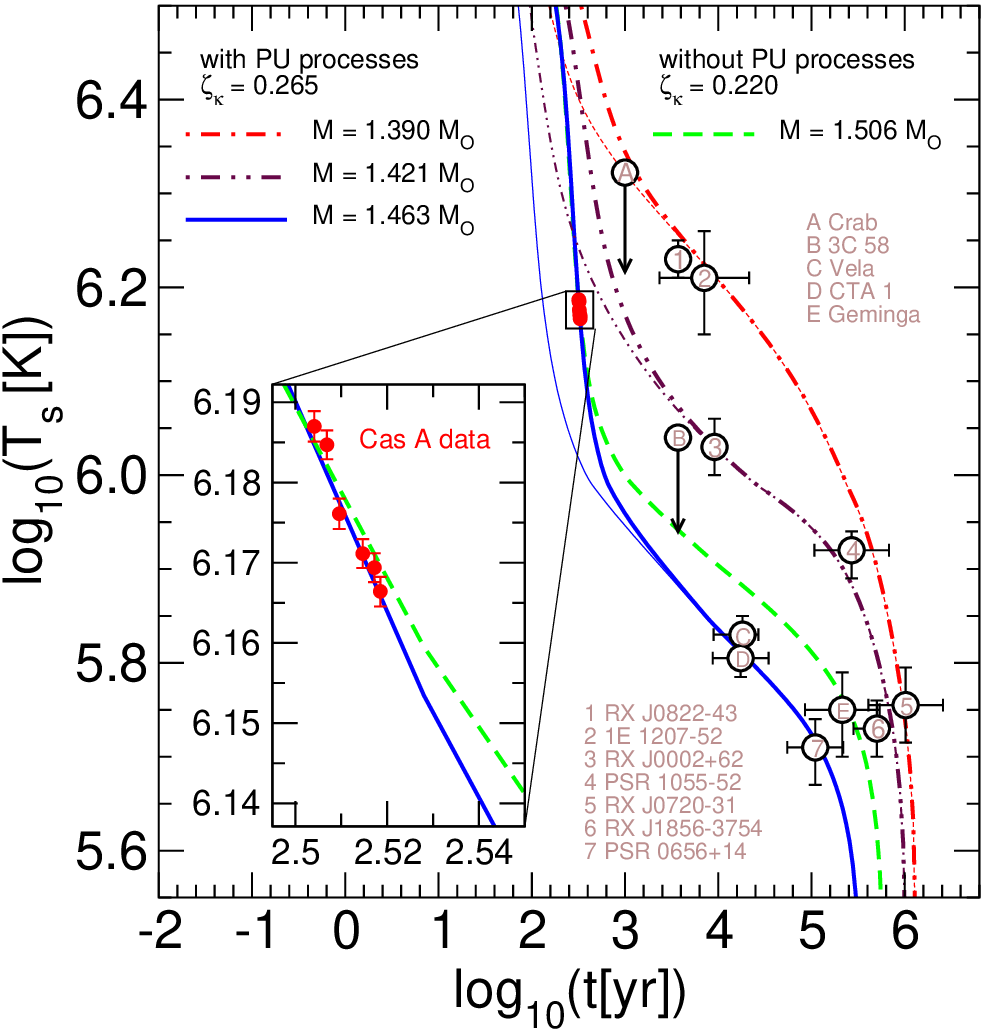}
   \caption{(Color online) Cooling of NSs
     with 
     nuclear medium effects, with and without pion condensation (PU),
      see also Fig.~17 of \cite{Blaschke:2004vq}.
     Data from Refs.~\cite{Page:2004fy,Yakovlev:2010ed}.}
   \label{Fig1}
\end{figure}

{\it The Neutron Star in Cas A.--}  
The ingredients of the nuclear medium cooling scenario discussed above lead
to neutron star cooling curves in Fig.~17 of Ref. \cite{Blaschke:2004vq} where 
model I for the proton gap has been used and the role of the heat conductivity 
on the hot early stages of hadronic neutron star cooling was elucidated.
In Fig.~\ref{Fig1} we redraw those cooling curves allowing for a minor
readjustment of
the heat conductivity parameter.
The bold curves are for a heat conductivity
  suppressed by a factor of $\zeta_\kappa = 0.265$, while the thin
  lines are for the unsuppressed heat conductivity of
  \cite{Baiko:2001cj}.
One sees that for a suppression factor of
$\zeta_\kappa = 0.265$ and a stellar mass of $M =1.463~{\rm M}_\odot$
(blue bold solid line) we are able to fit the temperature data for
Cas~A perfectly, as can be seen from the magnified 10-year epoch for
which high-precision cooling data exist.
This star is our best-fit model.
 Lowering the neutron star mass to $M =1.390~{\rm M}_\odot$ (red dash-dotted 
line) the whole set of available cooling data is covered.
 Assuming the absence of a pion condensate in the core of a neutron star,
  the Cas~A cooling data can still be reproduced by reducing $\zeta_{\kappa}$
  from 0.265 to 0.220 and 
   readjusting the neutron star mass to a somewhat
  higher value of $1.506~M_{\odot}$, see Fig.~1.
  The proton gap of model II is significantly smaller than that
   of model I. Nevertheless, the Cas~A data can still
    be nicely fitted for $\zeta_{\kappa} \leq 0.015$ and neutron star masses
    $M\ge 1.73~M_\odot$.
\begin{figure}[!tbh]
   \includegraphics[width=0.45\textwidth,height=0.25\textwidth]{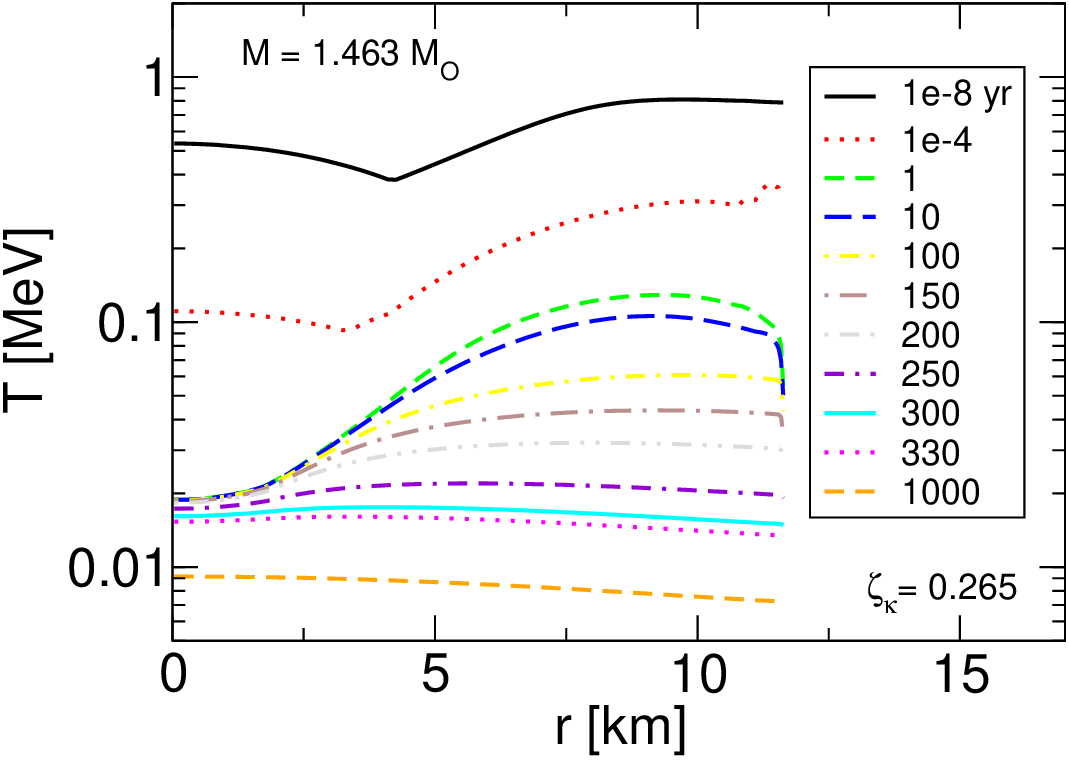}
   \caption{(Color online) Snapshots of temperature profiles for the
     Cas~A cooling curve (blue bold solid line) of Fig.~\ref{Fig1}.}
   \label{Fig2}
\end{figure}

To demonstrate the impact of the heat conductivity on the cooling
  process we present in Fig.~\ref{Fig2} the temperature profiles for
  the $1.463~M_{\odot}$ neutron star ($\zeta_{\kappa}=0.265$) for
  stellar ages from $10^{-8}$ to $10^3$ years.
One sees that the heat conductivity is important during the 
first $t\lsim 300$ years and would thus affect the cooling history 
of Cas~A.

\begin{figure}[!h]
  \includegraphics[width=0.45\textwidth,height=0.3\textwidth]{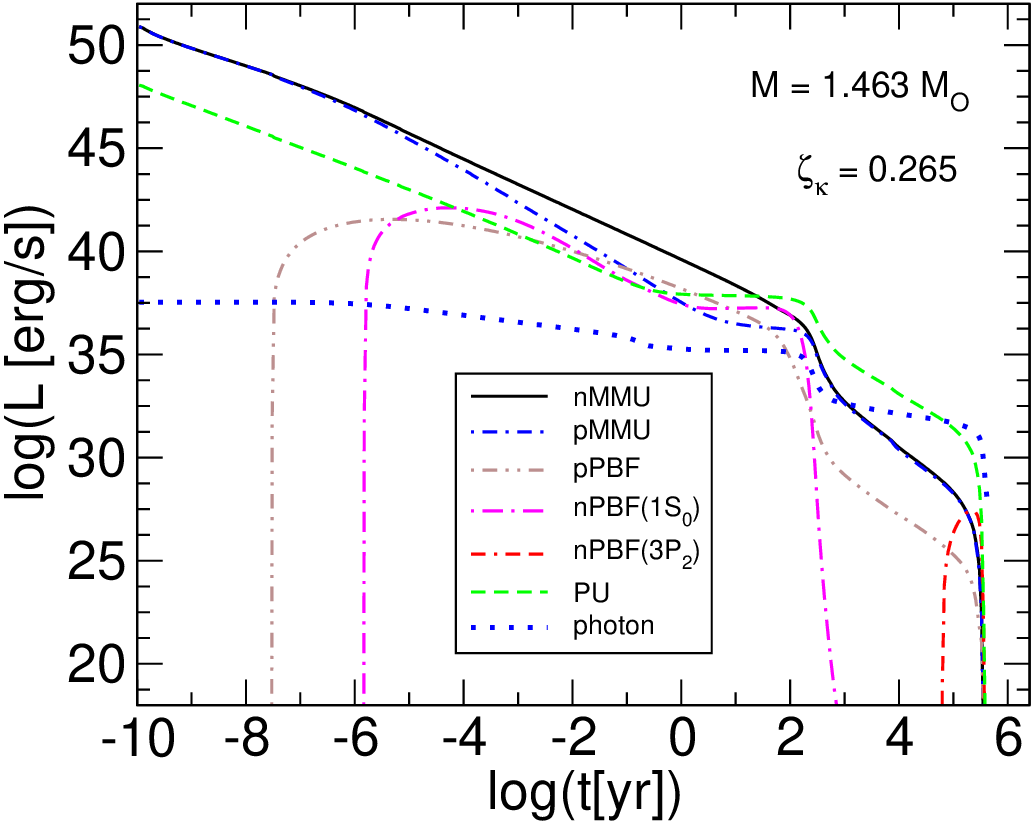}
  \caption{(Color online) Individual contributions of the
    cooling processes nMMU and pMMU, $1S_0$ pPBF and nPBF, $3P_2$
    nPBF, PU, and surface photon emission to the total stellar
    luminosity for the neutron star shown in Fig.~\ref{Fig2}.  }
   \label{Fig3}
\end{figure}
In Fig.~\ref{Fig3} we show the individual contributions of the cooling
processes of our scenario to the total neutron star luminosity for
the neutron star, $M = 1.463~M_{\odot}$ and
$\zeta_{\kappa}=0.265$, which best reproduces the cooling of Cas~A in
Fig.~\ref{Fig1}.  We see that the nMMU is the most efficient process
in our scenario, while all PBF processes are less important.
The MnB and MpB luminosities dominate over those of PBF.
They are not shown in Fig.~\ref{Fig3} since they have rather similar shapes
as the nMMU and pMMU curves.
Note that PU processes affect the NS cooling primarily at later times.

\medskip

{\it Summary and Conclusion.--}
We have shown in this Letter that the nuclear medium cooling scenario
allows one to nicely explain the observed rapid cooling of the neutron star 
in Cas~A. 
As demonstrated already in \cite{Blaschke:2004vq},
in this scenario the rapid cooling of very young objects like Cas~A 
is due to the efficient MMU and MnB processes, a very low (almost zero) value 
of the $^3P_2$ neutron gap, a large proton gap and a small thermal conductivity 
of neutron star matter.

Our explanation of the Cas~A cooling constitutes an alternative
  to that of \cite{Page:2010aw,Shternin:2010qi}, which is
  based on a strong PBF process due to $^3P_2$ superfluidity in
  neutron star interiors.  We support, however, the conclusion of
these authors that a large value of the proton gap is preferable,
albeit not necessarily in the entire neutron star core.  The results
presented in Fig.~\ref{Fig1} predict that the rapid cooling observed
for Cas~A will continue for a few  more decades until it slows
down when the temperature domain around $\mbox{log}_{10} T_s[K] =6 $
is reached.  Already in about ten years from now of continued
monitoring, the high accuracy of the data for Cas~A's surface
temperature will allow one to distinguish at the 2 $\sigma$
level between models with and without additional fast cooling
processes (pion condensation in our case).

To discriminate between alternative cooling scenarios further tests
may be considered, such as the comparison of log N-log S distributions
from population synthesis with the observed one for isolated neutron stars.
A recent study of this kind \cite{Popov:2004ey} favored
model II for the proton gaps.  Thus it may well be that actual values
of the thermal conductivity are smaller than assumed in Fig.~1, or
that there are other important aspects of the cooling of Cas~A
which have not yet been identified.

\begin{acknowledgments}
  We thank E.E. Kolomeitsev for discussions.  The work of H.G. was
  supported in part by the Volkswagen Foundation grant No 85 182 and
  by the Polish Ministry for Science and Higher Education (MNiSW) grant
  ``CompStar-POL''.
  D.B. was supported by MNiSW grants No. NN 202 23 1837,
  by the Russian Fund for Basic Research (RFBR) grant No. 11-02-01538-a
  and by CompStar, a research networking
  programme of the European Science Foundation.
  The work of D.N.V. was supported by the Alliance Program of the
  Helmholtz Association (HAS216/EMMI). F.W. is supported by the
  National Science Foundation (USA) under Grant PHY-0854699.
\end{acknowledgments}

\end{document}